\documentstyle[12pt, epsf]{article}    

\begin{document}
\input epsf

\newcommand{\beq}{\begin{equation}}
\newcommand{\eeq}{\end{equation}}
\newcommand{\bea}{\begin{eqnarray}}
\newcommand{\ba}{\begin{eqnarray*}}
\newcommand{\eea}{\end{eqnarray}}
\newcommand{\ea}{\end{eqnarray*}}
\newcommand{\tyhja}{\vspace{0.5cm}}
\newcommand{\mn}{_{\mu\nu}}
\newcommand{\al}{\left\langle}
\newcommand{\ar}{\right\rangle}
\renewcommand{\vec}[1]{{\bf#1}}
\renewcommand{\Re}{\mathop{\hbox{Re}}}
\renewcommand{\Im}{\mathop{\hbox{Im}}}
\newcommand{\simleq}{\mathop{\mbox{{\tiny ${\stackrel{\textstyle <}{\sim}}$}}}}
\newcommand{\simgeq}{\mathop{\mbox{{\tiny ${\stackrel{\textstyle >}{\sim}}$}}}}
\newcommand{\doo}{\partial}

\begin{titlepage}
\begin{flushright}
HU-TFT-96-27\\
hep-lat/9607023\\
July 8, 1996
\end{flushright}
\begin{centering}
\vfill
{\Large\bf Dimensionally reduced U(1)+Higgs theory in the broken phase}
\vspace{1cm}

Mika Karjalainen$^1$ and Janne Peisa$^2$
\vspace{0.5cm}

{\em Department of Physics \\ P.O. Box 9 \\ 
FIN-00014 University of Helsinki, Finland}
\vspace{2cm}

{\bf Abstract}

\vspace{0.5cm}
\end{centering}

\noindent
We apply dimensional reduction to the finite temperature U(1)+Higgs theory
and study the properties of the reduced 3-dimensional theory in the broken
phase using lattice Monte Carlo simulations. We compute analytically
the scalar condensate in optimized 2-loop perturbation theory and the
correlators in 1-loop perturbation theory. These quantities are also
calculated numerically. The two results for the condensate
agree well but a 25\% difference is observed for the scalar correlator, 
indicating the need for optimized 2-loop perturbative results.
\vfill

\footnotetext[1]{Mika.Karjalainen@helsinki.fi; 
$^{2}$Janne.Peisa@helsinki.fi}
\end{titlepage}

\section{Introduction}  

The U(1)+Higgs theory has been studied extensively in four dimensions
\cite{4d,dimred, frame}, but only to a limited extent in 3d \cite{3d}. The 
3d studies are mainly motivated by the fact that the theory
then is directly the Ginzburg-Landau model of superconductivity.
That the 3d theory also arises as
an effective theory of the finite $T$ 4d
U(1)+Higgs theory 
is a strong additional motivation to study
it. Although the really physical
case is the SU(2)+Higgs theory, it is nevertheless of great interest
to investigate how changing the group from SU(2) to U(1) changes the
dynamics. This may also help in understanding the phenomena in the
full SU(2)$\times$U(1) standard model. The purpose of this article is to study
the 3d U(1)+Higgs theory, mainly in the broken phase
and for values of the couplings which are appropriate when the
3d theory is an effective theory of the finite $T$ 4d
U(1)+Higgs theory \cite{dimred, frame}.

Perturbation theory in U(1)+Higgs model breaks down at the tree-level critical
temperature, as the expansion parameter diverges when $T \to T_c$.
Although the lattice $T_c$ differs slightly from the perturbative one, these 
temperatures are so close that perturbation theory is not valid 
at the lattice $T_c$ either. 
At temperatures below $T_c$, but still higher than all the 
mass scales in the theory, the expansion parameter
is small and one expects that perturbation theory works rather well there. 
The main goal of this article is to
make statements on the accuracy of 
perturbation theory at intermediate temperatures, where the expansion
parameter is not too large and also try to estimate the temperature, at
which perturbation theory is not valid anymore. To this end we compute
on the one hand analytically in perturbation  
theory and on the other hand numerically with lattice Monte Carlo
techniques the scalar condensate and the scalar and vector correlators.
A comparison of the results gives a quantitative assessment of the
accuracy of perturbation theory. We shall in a later paper
\cite{kklp} study the phase transition properties of the theory.

The accuracy of perturbation theory depends not only on the number
of the loops but also on the optimization method used. In this paper
we compare ordinary $\hbar$-expansion with another optimized method,
namely the CW-method. The latter can also be applied
when the broken minimum is radiatively generated, the first one requires
the existence of the broken minimum already in the tree potential.

First we use the dimensional reduction to obtain an effective model
for 4d finite temperature U(1)+Higgs theory. The idea of dimensional
reduction is to integrate out all the non-zero
Matsubara modes which are heavy at high temperature \cite{dr, generic}.
The effective 3d theory obtained after reduction provides an 
excellent approximation
of the original theory \cite{frame} when two basic conditions are 
satisfied: the coupling constants are small,
so that we can relate the parameters of the original 4d and the effective 
3d models pertubatively and $T$ is larger than all the relevant mass scales 
in the theory \cite{ae,la}. 
Dimensional reduction has one major advantage over direct finite
temperature 4d simulations. In particular the dimensionally
reduced 3d theory contains less mass scales than the original theory
(if we integrate over the temporal component of the gauge-field, there
are two mass scales less), resulting in a huge gain in computer
simulations. Because all fermions are integrated out at the reduction
process, one could also study the theories with chiral fermions, which
are problematic on the lattice in 4d.

To be able to compare our non-perturbative Monte Carlo
data with perturbative results, we have to determine relevant
observables accurately. An essential ingredient when comparing the
perturbation theory to lattice calculations are the continuum-lattice
mapping formulae, which are exact in 3d \cite{ct}. Thus the continuum
limit can be carried out under controlled conditions.

\section{Dimensional reduction}

The Lagrangian of the 4d U(1)+Higgs theory at finite temperature 
without fermions is in standard notation
\beq
	{\cal L} = \frac{1}{4}F_{\mu\nu}F_{\mu\nu} +
\left(D_{\mu}\phi\right)^{\ast} \left(D_{\mu}\phi\right) +
m^2\phi^{\ast}\phi +
\lambda\left(\phi^{\ast}\phi\right)^2. \label{L} 
\eeq
One could also include fermions as this results in the
same final effective 3d theory but with different parameter values. 

In the following we shall choose 
the four dimensional coupling $g=1/3$ and the 4d tree masses 
$m_H=-2 m^2=30, 35, 60$ GeV and $m_W= m_H \sqrt{g/2\lambda} =  80.6$ GeV.
We call the arbitary mass parameter $m_H$ the tree-level Higgs
mass. The choice of these values of 
$m_H$ is motivated by the fact that for smaller values the transition
becomes too strong for dimensional reduction to be accurate and for
larger values perturbation theory becomes increasingly unreliable

The dimensional reduction of U(1)+Higgs theory is presented in
\cite{dimred} and we just collect the results here. After integrating
over the scale $\pi T$ the typical scale of remaining 
3d theory is $\sim g^2T$. There is still an adjoint Higgs field $A_0$ present
in the theory, but its mass $m_{A_0} \sim g T$ is large compared with
the scale of the theory. Hence also $A_0$ can be integrated out
perturbatively.  
The action which is obtained after eliminating all heavy
degrees of freedom is
\begin{equation}
S_{\mbox{\scriptsize eff}}\left[A_i({\bf x}),\phi_3({\bf x})\right] = \int d^3x 
\bigg[\frac{1}{4}F_{ij}F_{ij} + (D_i\phi_3)^2 + m_3^2 \phi_3^{\ast}\phi_3 +
\lambda_3 \left(\phi_3^{\ast}\phi_3\right)^2\bigg].
\label{L3}
\end{equation}
The relations between 3d and 4d parameters are: 
\beq
	g_3^2=g^2(\mu)T,
\eeq
\beq
	\lambda_3=\lambda(\mu)T-\frac{g_3^4}{8\pi m_D},
\eeq
\bea
m_3^2(\mu_3) & = &
\bigg[\frac{1}{4}g_3^2T+\frac{1}{3}\lambda_3T+\frac{g_3^2}{16\pi^2}
\left(-\frac{8}{9}g_3^2+\frac{2}{3}\lambda_3\right)\bigg] \nonumber \\
& & -\frac{1}{2}m_H^2 + {f_{2m} \over 16\pi^2}\left(\log {3T \over
\mu_3} +c\right) \nonumber \\ & & -\frac{g_3^2m_D}{4\pi}-\frac{g_3^4}{8\pi^2}\left(\log
\frac{\mu_3}{2m_D} +\frac{1}{2}\right), \label{m32} 
\eea
where we have
\beq
	m_D^2=\frac{1}{3}g^2(\mu)T^2,
\eeq
\beq
	f_{2m} = -4 g_3^4 + 8\lambda_3 g_3^2 -8\lambda_3^2. 
\eeq
The constant $c=-0.348725$.

The mean field dependent masses are
\bea
	m_T=g_3\phi, & m_1^2 = m_3^2(\mu_3)+3\lambda_3\phi^2, &
m_2^2=m_3^2(\mu_3)+\lambda_3\phi^2. \label{masses} 
\eea

The couplings $g_3^2$ and
$\lambda_3$ are 3d renormalization group
invariant. 
When propagators are calculated perturbatively we notice that
the mass parameter $m_3^2$ contains both linear and logaritmic divergences.
The appearance of a logarithmic divergence introduces
a scale $\mu_3$ into our theory. 
We can choose this extra scale to be
the scale parameter of dimensional regularization in 3d $\overline{\rm MS}$ 
scheme. The tree parameters must be transformed with
renormalization group equations when $\mu_3$ varies if we want to keep
the physical content of the theory unchanged.

All the information of the 4d theory is contained in just
three parameters: a dimensionful gauge coupling constant $g_3^2$, which defines
the typical scale of the system and two dimensionless numbers which we
define as  
\beq
 x \equiv \frac{\lambda_3}{g_3^2}, \,\,\,\, y \equiv
\frac{m_3^2(g_3^2)}{g_3^4}. 
\eeq
The parameter $x$ is essentially proportional to the ratio of the squares 
of the
scalar and vector masses in the broken phase, while $y$ is related
to the temperature, $y \sim (T-T_c)/T_c$, so that at tree level $y=0$
describes the phase transition point. In terms of 4d parameters they can be
expressed as
\beq
	x=\frac{1}{2}\frac{m_H^2}{m_W^2}-\frac{\sqrt{3} g}{8 \pi},
	\label{aks}
\eeq
\bea
	y &=& \frac{1}{4g^2}+\frac{1}{3}\left(x+\frac{\sqrt{3} g}{8 \pi}\right)
	\frac{1}{g^2} + \frac{1}{16 \pi^2}\left[-\frac{8}{9}
	+\frac{2}{3}\left(x+\frac{\sqrt{3} g}{8 \pi}\right)\right]
	\nonumber \\ 
	& & -\frac{1}{4\pi\sqrt{3}g} - \frac{1}{8\pi^2}\left(
	\log \frac{3\sqrt{3}}{2g}+c+\frac{1}{2}\right)-\frac{m_H^2}{2g^4T^2}
	\nonumber \\
	& & + \frac{1}{16 \pi^2}\left(-4 +8x-8x^2\right)\left(\log 
	\frac{3}{g^2}+c\right). \label{yy}
\eea
The 4d tree level Higgs mass values $m_H=30, 35$ and $60$ GeV
correspond to $x=0.0463, 0.0713$ and $x=0.254$ according to equation
(\ref{aks}).  

\section{Perturbative calculation of the properties of the system in the
broken phase}

\subsection{3d effective potential}

\begin{figure}[t]
  \vspace*{-4cm}
  \hspace*{-3.75cm}
  \epsfysize=8cm
  \epsffile[0 550 300 800]{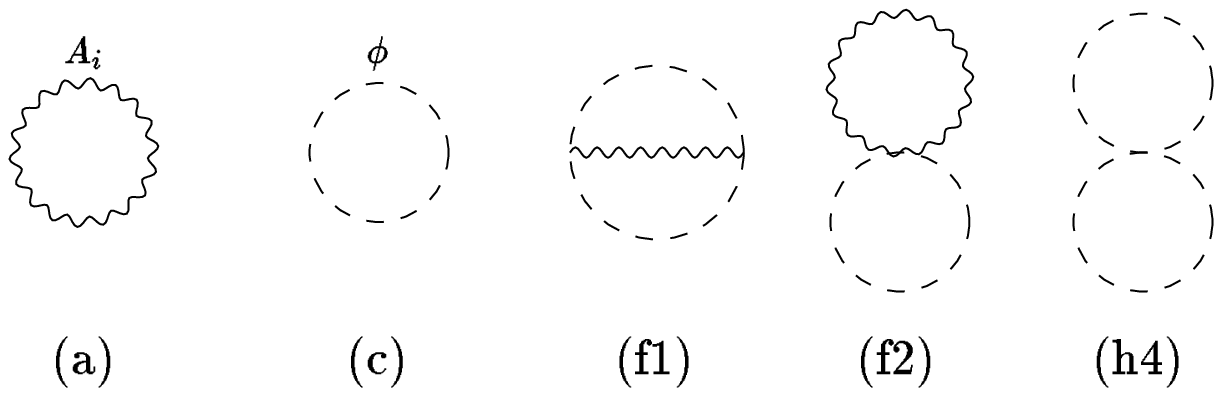}
  \vspace*{-1.0cm}
  \caption{The 2-loop diagrams contributing to the 3d effective potential.}
  \label{2l}
\end{figure}

The tree potential is obtained from the Lagrangian
(\ref{L3})
\[
V_0 = {1\over 2} m_3^2 \phi^2 + {1\over 4} \lambda_3
\phi^4,
\]
and the 1-loop part of the effective potential in 3d is  
\beq
	V_1 = -\frac{1}{12\pi}\left[2m_T^3+m_1^3+m_2^3\right]. \label{V1}
\eeq
The 2-loop potential in 3d contains diagrams in Fig.~\ref{2l} ($\times
1/(16\pi^2)$) 
\bea
	(a) & = & -g^4\phi^2 D_{VVS}(m_T,m_T,m_1), \nonumber \\
	(c) & = & -\lambda^2\phi^2 \left[3
H(m_1,m_1,m_1)+H(m_1,m_2,m_2)\right], \nonumber \\ 
	(f1) & = & -\frac{1}{2}g^2 D_{SSV}(m_1,m_2,m_T), \nonumber \\ 
	(f2) & = & g^2m_T(m_1+m_2), \nonumber \\
	(h4) & = &
\frac{1}{4}\lambda\left(3m_1^2+2m_1m_2+3m_2^2\right), \label{2} 
\eea
where we have used the results and diagram notations of \cite{dimred}
and 
\bea
\hspace*{-0.5cm}
	D_{SSV}(m_1,m_2,M) &=&
	\left(M^2-2m_1^2-2m_2^2\right)H(M,m_1,m_2) \nonumber \\ 
	& & + \frac{\left( m_1^2-m_2^2 \right)^2}{M^2}[H(M,m_1,m_2)
	-H(0,m_1,m_2)]  \nonumber \\ 
	& & + \frac{1}{M}{(m_1+m_2)\left[M^2+(m_1-m_2)^2\right]-M m_1 m_2},
	\label{SSV} 
\eea
\bea
	D_{VVS}(M,M,m) &=& 
	\bigg(2H(M,M,m)-\frac{1}{2}H(M,m,0)
	+ \frac{m^2}{M^2}[H(M,m,0) \nonumber \\  
	& & -H(M,M,m)] + \frac{m^4}{4M^4}[H(m,0,0)+H(M,M,m) \nonumber \\  
	& & -2H(M,m,0)] -{m \over 2M^{}}-{m^2 \over 4M^2}\bigg), \label{VVS}
\eea
\beq
  H(m_1,m_2,m_3) = \log \frac{\mu_3}{m_1+m_2+m_3} + \frac{1}{2}.
\eeq

\subsection{Choosing the scale \label{scale}}

As discussed previously the only scale dependent parameter in this theory
is the mass parameter $m_3^2(\mu_3)$, as the coupling constants do not
run. We also know that the effective potential is scale independent to
a certain order. On the other hand for example the tree as well as the 
2-loop effective potentials on their own are explicitly $\mu_3$ dependent, 
the scale dependence cancels to this order
when they are combined. The question is now, how to choose the scale
so that the higher order loop corrections are small and predictions of physical
parameters can be made. 

In many theories, including this, the problem involves large
logarithms. The 2-loop effective potential, which carries the
logarithms is only weakly $\mu_3$ dependent. One should choose
the scale so that the effect of these logarithmic terms is minimized
and it is quite natural to achieve this by demanding that the 2-loop
contribution vanishes totally. This is not exactly sufficient because 
the effective potential is not renormalization
group invariant, it is not a physical quantity. Only the difference
$V(\phi_1)-V(\phi_2)$ is measurable. One should then really look at
the derivative $dV/d\phi$, which is $\mu_3$ independent. 

\begin{figure}
  \centering
  \leavevmode
  \epsfysize=8cm
  \epsfbox{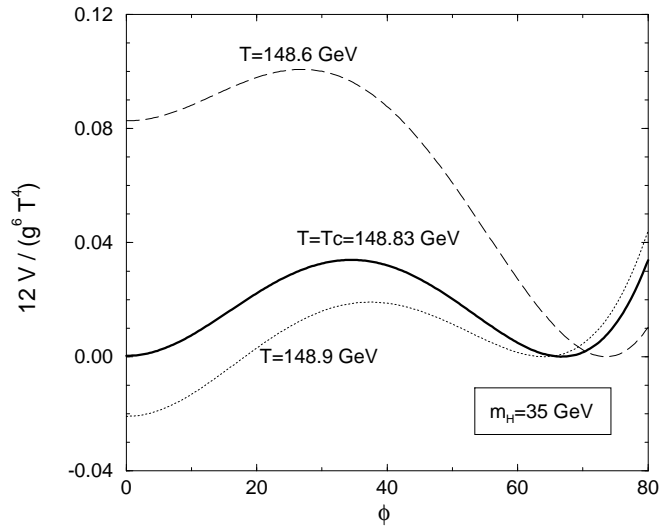}
  \caption{The renormalization group improved 2-loop potential 
  near the phase transition point, with $m_H=35$ GeV.}
  \label{VmH35}
\end{figure}

\begin{figure}
  \centering
  \leavevmode
  \epsfysize=8cm
  \epsfbox{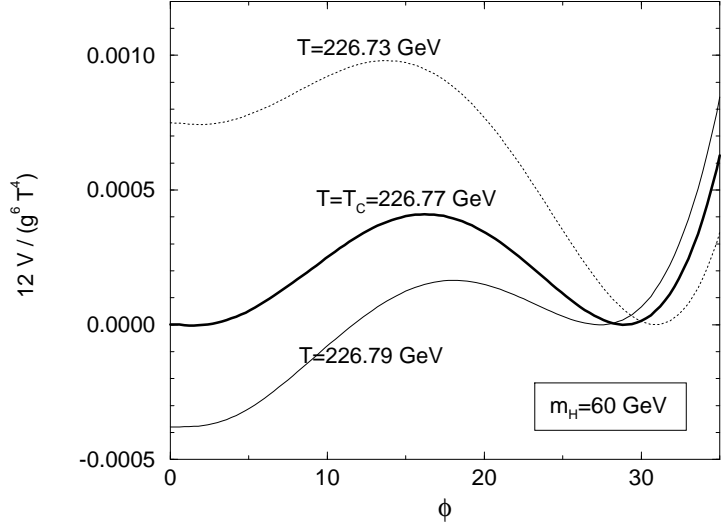}
  \caption{The 2-loop potential with $m_H=60$ GeV.}
  \label{VmH60}
\end{figure}

\begin{figure}
  \centering
  \leavevmode
  \epsfysize=8cm
  \epsfbox{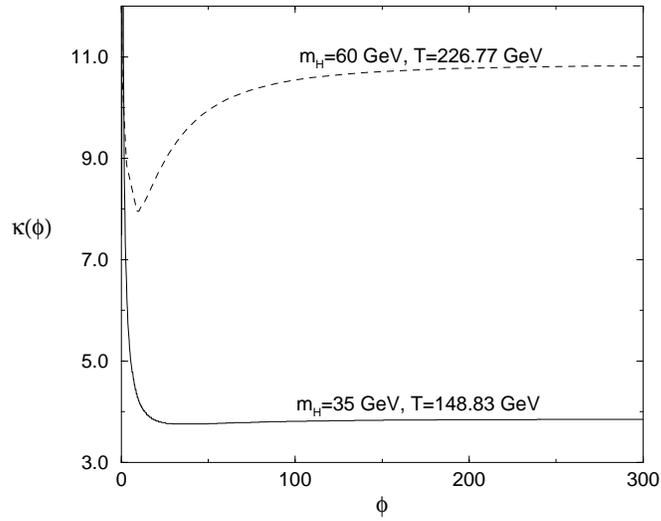}
  \caption{Fitted $\kappa(\phi)$:s with $m_H=35$ GeV, $T=148.83$ GeV $\approx T_c$ and $m_H=60$ GeV, $T=226.77$ GeV $\approx T_c$.}
  \label{kappas}
\end{figure}

In the effective U(1)+Higgs theory we have three different mass scales
($m_T, m_1, m_2$) and the determination of proper $\mu_3$ is
nontrivial. Let's first assume that $m_T \gg m_1, m_2$ so the Higgs
mass is so small that we can neglect the $m_1$ and $m_2$
terms. Solving then $dV_2=0$ we find that the choice of $\mu_3$ which minimizes
the 2-loop potential is 
\beq
	\mu_3 = \kappa m_T, \label{kappa}
\eeq
where $\kappa$ has to be
determined using some other criterion of minimizing the effect of
2-loop terms. Here we demand that the value of $\phi$ at the
critical temperature $T_c$ is the same for the 1-loop and 2-loop
potentials in the broken phase. This parametrization with constant
$\kappa$ works rather well, when the
Higgs mass is small. When all three mass scales ($m_T, m_1, m_2$) are 
included, we generalize (\ref{kappa}) by choosing 
\beq
	\mu_3(\phi) = \kappa(\phi) m_T,
\eeq
where $\kappa(\phi)$ is fitted numerically at every point $\phi$ so that
the derivative of the 2-loop contribution vanishes. Pictures
of potentials with different Higgs masses are presented in
Figures~\ref{VmH35}~and~\ref{VmH60}. 
The corresponding $\kappa(\phi)$:s can be seen in Figure~\ref{kappas}. 
Note that the value of $\kappa(\phi)$ and therefore also $\mu_3(\phi)$ is not
well determined for very small values of $\phi$ because
perturbation theory is not valid there.  

\subsection{The ground state energy and the scalar condensate in the
broken phase} 

We will now calculate analytically the ground state energy of the theory 
and the scalar condensate using methods described in \cite{frame, nonpert}.

In perturbation theory the effective potential is derived as an
expansion where the expansion parameter is the Planck constant $\hbar$
\beq
	V(\phi) = \sum_{n=0}^{N} \hbar^{n}V_n(\phi).    \label{V}
\eeq
Although the potential is gauge dependent, its value in the broken
minimum, which  
gives the vacuum expectation value $\nu(T)$ of $\phi$ is gauge
independent and determined by  
\beq
	\left. \sum_{n=0}^{N} \hbar^{n}         \frac{dV_n(\phi)}{d\phi}\right|_{\phi=\nu(T)}=0. \label{nu}
\eeq
The aim is now to solve $\nu(T)$ from (\ref{nu}).  

We look at the so called "classical" regime, where
spontaneous symmetry breaking occurs already on the tree-level of the
potential. This means that there is a (non-trivial) solution
$\nu_0^2(T)=-m_3^2/\lambda_3$ so that 
\beq
	\left. \frac{dV_0(\phi)}{d \phi}\right|_{\phi=\nu_0(T)}=0,
\eeq
which gives the leading approximation to $V_{min}$
\beq
	V_{min} \sim V_0(\nu_0)= -\frac{m^4_3}{4 \lambda_3}.
\eeq
If perturbation theory works well we can use this saddle point value
as a good start ($\nu(T)=\nu_0(T)+{\cal O}(\hbar)$) and solve (\ref{nu})
perturbatively 
\beq
	\varphi=\sum_{n=0}^N \hbar^n \varphi_n,    \label{fi}
\eeq
where we have denoted $\varphi = \nu^2(T)$ and $\varphi_0 = \nu_0^2(T)$. 

Perturbation theory works only when the expansion parameter is
small enough. When we look at 1-loop and 2-loop
potentials separately we notice that typical 1-loop terms have
$m_T^2/2\pi$ with them and 2-loop terms include $g_3^2
m_T^2/(2\pi)^2$. The expansion parameter $\rho$ is the ratio of these 
\beq
	\rho=\frac{g_3^2}{2\pi m_T}=\sqrt{\frac{g_3^2 \lambda_3}{-m_3^2 \pi}}. \label{rho}
\eeq
So the "classical" regime approximation is valid when
\bea
	m_3^2<0, & |m_3^2| \gg g_3^2 \lambda_3.
\eea

Using (\ref{V}) and (\ref{fi}) we have at the minimum
\bea
	V'_0(\varphi_0+\hbar \varphi_1+\hbar^2 \varphi_2)  & & \nonumber \\ 
	+\hbar V'_1(\varphi_0+\hbar \varphi_1) + \hbar^2 V'_1(\varphi_0) 
	+ {\cal O}(\hbar^3)& = & 0.
\eea
Here primes mean that we take partial derivatives with respect to
$\phi^2$ and evaluate the expression at the value written in
parenthesis. When we also 
expand all $V_n$:s near $\varphi_0$ we get
\bea
	V'_0(\varphi_0)+ V''_0(\varphi_0)[\hbar \varphi_1+\hbar^2
\varphi_2] & & \nonumber \\ + \frac{1}{2} V'''_0(\varphi_0)[\hbar^2
\varphi^2_1] +  
\hbar V'_1(\varphi_0) & & \nonumber \\ + \hbar V''_1(\varphi_0)[\hbar
\varphi_1] + \hbar^2 V'_2(\varphi_0) + {\cal O}(\hbar^3) & = & 0. 
\eea
All degrees of $\hbar$ must vanish separately, which gives 
\beq
  \varphi_1 = -\frac{V'_1(\varphi_0)}{V''_0(\varphi_0)}. \label{fi1} 
\eeq
From now on we will not write down the $\varphi_0$-term any more,
although all potential terms are still evaluated at this saddle
point. 

The value of the effective potential at the minimum up to 3 loops is
calculated by expanding (\ref{V}) with (\ref{fi}) 
\bea
	V_{min}& = & V_0 + \hbar V_1 + \hbar^2 \left[V_2 + \frac{1}{2}
V'_1 \varphi_1 \right] \nonumber \\ & & + \hbar^3 \left[V_3 + V'_2
\varphi_1 + \frac{1}{2} V''_1 \varphi^2_1 \right] + {\cal O}(\hbar^4),
\label{Vmin} 
\eea
which gives the ground state energy in the broken phase. 
With this, we can easily calculate the scalar condensate
$\langle\phi^{\ast}\phi(\mu_3)\rangle$ in the broken phase 
\beq
	\langle\phi^{\ast}\phi(\mu_3)\rangle = \frac{d
V_{min}}{dm_3^2(\mu_3)}. \label{cond}
\eeq

Although the 3d-effective potential is calculated only up to 2-loops,
we can make some predictions of the 3-loop potential term. 

There is a 3-loop order logaritmic $\mu_3$-dependence in the 1-loop
potential $V_1$ because of the $\mu_3$-dependence of $m_1(\phi)$. Because the
3-loop level effective potential is $\mu_3$-independent, there must be a
term in $V_3$ which cancels this. So, $V_3$ must include a term 
\bea
	V_{3a}& = &-\mu_3 \frac{\doo V_1}{\doo \mu_3}\left[\log
\frac{\mu_3}{m_T}+\frac{1}{2}\right] \nonumber \\ & = & \frac{\doo V_1}{\doo
m_3^2}\frac{f_{2m}}{16\pi^2}\left[\log
\frac{\mu_3}{m_T}+\frac{1}{2}\right],
\eea
and a $\mu_3$-dependence of $V_1 +V_3$ is of order
$\hbar^5$. When we calculate the ground state energy the last term of
(\ref{Vmin}) gives a term proportional to $1/m_2$. At the saddle point
$m_2 \rightarrow 0$ and this term diverges. To avoid
this, there must be another compensating term in $V_3$. Calculation 
of $V'_2 \varphi_2 + \frac{1}{2} V''_1
\varphi^2_1$ gives us this term 
\beq
	V_{3b}=\frac{1}{32}\left(4+3r^3\right)^2
\frac{g_3^4}{(4\pi)^3} \frac{m^2_T}{m_2}, 
\eeq
where 
\beq
	r^2 \equiv \frac{2\lambda_3}{g_3^2}.
\eeq
There will be other contributions also, which for pure scalar theory have been
explicitly computed by \cite{rajantie}. In SU(2)+Higgs theory this unknown 
information was included under one linear term \cite{dimred}, which can
be derived just on dimensional grounds. This term depends on masses,
$\phi$ and a dimensionless constant parameter $\beta$
\beq
	V_{3c}=\rho \frac{m^2_T}{2\pi}\beta=\frac{g_3^4 m_T}{(4\pi)^3}\beta.
\eeq
In U(1)+Higgs theory this kind of linear term is absent \cite{h} so the
unknown part cannot be found using this parametrization. 

The known part of the 3-loop potential $V_3$ can be written in form 
\bea
	V_3 & = & V_{3a}+V_{3b} \nonumber \\ & = & \frac{\doo
V_1}{\doo m_3^2}\frac{f_{2m}}{16\pi^2}\left[\log
\frac{\mu_3}{m_T}+\frac{1}{2}\right]+\frac{1}{32}\left(4+3r^2\right)^2
\frac{g_3^4}{(4\pi)^3} \frac{m^2_T}{m_2}. \label{V3} 
\eea

To be able to calculate the scalar condensate we must now find
$V_{min}$ explicitly. First we have to calculate $\varphi_0$ and
$\varphi_1$. Using 
(\ref{fi1}) with (\ref{V1}) and (\ref{2}), we get  
\bea
	\varphi_0 & = & \frac{-m_3^2(\mu_3)}{\lambda_3} \\
	\varphi_1 & = &\frac{1}{4\pi r^2}\left(4m_T + 3 m_1 r^2 +m_2
r^2\right)={3 \over 4\pi}\left({4 \over r^2} + r\right)m_T, 
\eea
where masses are evaluated at the saddle point
\beq
	m_1=r m_T=\sqrt{-2 m_3^2(\mu_3)}, \hspace*{0.5cm} m_2=0,
\eeq
and
\beq
	\frac{d m_T}{dm_3^2}=-\frac{1}{r^2 m_T}, \hspace*{0.5cm}
\frac{d m_1}{d m_3^2} = -\frac{1}{m_1}. 
\eeq
Inserting these into the potentials we get
\bea
	V_0 & = & -\frac{m_3^4(\mu_3)}{\lambda_3}, \\ \vspace*{0.3cm}
	V_1 & = & -\frac{1}{12\pi}\left(2 + r^2 \right)m_T, \\
\vspace*{0.3cm} 
	V_2 + \frac{1}{2}V'_1 \varphi_1 & = & \frac{1}{16\pi^2}g_3^2
m_T^2 \left[f_1(r)\left(\log \frac{\mu_3}{m_T} +\frac{1}{2}\right) +
q_1(r)\right], 
\eea
where
\bea
	f_1(r) & = & -\left(2 - r^2 + r^4\right), \\ \vspace*{0.3cm}
	q_1(r) & = & \left(2 -r^2 +\frac{1}{4}r^4 \right)\log (2 +r)
-2r +\frac{1}{4}r^4 \nonumber \\ & & -\frac{1}{2}r^3 + \frac{3}{4}r^4
\log (3r) -\frac{2}{r^2}-\frac{3}{4}r^4. 
\eea
Now a lengthy computation shows that
\bea
	V'_2 \varphi_1 + \frac{1}{2}V''_1 \varphi_1^2 & = &
\frac{3}{2}\left( r + \frac{4}{3r^2} \right) \frac{m_T
f_{2m}}{(4\pi)^3} \left( \log \frac{\mu_3}{m_T} + \frac{1}{2}\right)
\nonumber \\ & & + \beta_{disc}(r)\frac{m_T g_3^4}{(4\pi)^3} \nonumber
\\ & & + \frac{1}{32}\left(4 + 3r^3\right)^2 \frac{g_3^4}{(4\pi)^3}
\frac{m_T^2}{m_2}, 
\eea
where the divergent $m_2$ part cancels with $V_3$ and 
\bea
	\beta_{disc}(r) & = & {4+3r^3 \over 32r^4(2+r)} \bigg[ -64 -32
r +64 r^2 + 8 r^3 -36 r^4 -40 r^5 +10 r^6 \nonumber \\ & &  + 13 r^7 +
(48 r^6  +24 r^7) \log (3) + (64 r^6 + 32
r^7) \log (r)\nonumber \\ & &  - (32 r^4  + 16 r^5  +
32 r^6 + 16 r^7) \log(1+r) \nonumber \\ & & + (128 r^2 +64 r^3 - 96 r^4 
 -48 r^5 +32 r^6  + 16 r^7) \log(2+r) \bigg]. 
\eea
Using (\ref{V3}) and inserting these into (\ref{Vmin}) gives us the vacuum
energy density in the broken phase up to 3-loops 
\bea
	V_{min} & = & -\frac{m_3^4(\mu_3)}{\lambda_3}
-\frac{1}{12\pi}\left(2 + r^2 \right)m_T \nonumber \\ & & +
\frac{1}{16\pi^2}g_3^2 m_T^2 \left[f_1(r)\left(\log \frac{\mu_3}{m_T}
+\frac{1}{2}\right) + q_1(r)\right] \nonumber \\ & & +\left(r +
\frac{2}{r^2}\right){m_T f_{2m} \over (4\pi)^3}\left( \log
\frac{\mu_3}{m_T} + \frac{1}{2}\right) \nonumber \\ & & + {m_T g_3^4
\over (4\pi)^3} \beta_{disc}(r). 
\eea
To be able to compare the perturbative results with Monte Carlo data
we must calculate the analytic form of scalar condensate
$\langle\phi^{\ast}\phi\rangle$. This is easily done using
(\ref{cond}). 
In terms of previously defined parameters $x$ and $y$
the condensate is
\bea
	{\langle\phi^{\ast}\phi(\mu_3)\rangle \over g_3^2} & = &
{-y(\mu_3) \over 2x} +{1 \over 4\pi}\left(r +{2 \over r^2}\right)m_T
\nonumber \\ & & -{1 \over x}{1 \over 16\pi^2} \left[ f_1(r)\log
\frac{\mu_3}{m_T} + q_1(r)\right] \nonumber \\ & & -{ g_3^2 \over
(4\pi)^3}{1 \over r^2 m_T}\bigg[{f_{2m} \over g_3^4}\left(r+ {2 \over
r^2}\right) \left(\log {\mu_3 \over m_T} -{1 \over 2}\right) \nonumber
\\ & & +\beta_{disc}(r)\bigg], \label{co} 
\eea
where $m_T$ is 
\beq
	m_T = g_3^2 \sqrt{\frac{-y(\mu)}{x}}
\eeq
and
\beq
        y(\mu)=y +\frac{1}{16 \pi^2}\left(-4 +8x -8x^2\right) 
	\log\frac{g_3^2}{\mu}.
\eeq

\subsection{Coleman-Weinberg method}

The direct $\hbar$ expansion discussed above for calculating the 
scalar condensate is not valid at the phase transition point 
because the expansion parameter $\rho$ diverges. It also 
has higher order $\mu_3$-dependence, which is significant near the 
critical temperature. We can mainly avoid these problems using another, 
so called Coleman-Weinberg (CW) method which is much less scale dependent 
and can be used even beyond the critical temperature. With the CW method one
numerically determines the location of the broken minimum $\varphi_0(T)$ of
the renormalization group improved 2-loop effective potential and then 
calculates the ground state energy of the system. Finally one computes the
condensate using equation (\ref{cond}). Because of the singularities at the 
classical tree-level minimum ($V_3 \sim 1/m_2$ and $m_2 \to 0$), the 
ground state energy computed by CW-method differs from the result 
of the straightforward $\hbar$-expansion. One should then redefine the 2-loop
potential by including all leading singularities in it (Appendix B in 
\cite{nonpert}). This should not be a major problem in this case, 
because the difference is of order $\hbar^{\frac{5}{2}}$ and we have the 
effective potential only up to $\hbar^2$ as there are still unknown parts in 
the $\hbar^3$-level.  

\begin{table}
  \centering
  \begin{tabular}{|c|c|c|c|c|c|}
    \hline 
    $T$ & 106.4 & 115.8 & 125.4 & 126.8 & 127.5  \\
    \hline
    $\langle\phi^{\ast}\phi\rangle/T$ 
        & 2.395(6) & 1.551(8) & 0.807(4) & 0.705(11) & 0.638(5) \\
    \hline
    $\hbar$-exp. 
        & 2.369(6) & 1.515(9) & 0.707(20) &  0.573(27) & 0.495(37)\\
    \hline
    CW 
        & 2.392(2) & 1.550(1) & 0.784(2) & 0.672(2) & 0.614(2) \\
    \hline
  \end{tabular}
  \caption{Measured and computed values of
$\langle\phi^{\ast}\phi\rangle/T$ for $m_H=30$ GeV.  The first row
gives the lattice value extrapolated to continuum limit. The last two
rows show the 2-loop perturbative values using both $\hbar$ expansion
and CW-optimization. The errors on the perturbative
values are obtained by varying the scale parameter $\mu$ as indicated
in the text. \label{tbl} }
\end{table}

The results with both of these methods are presented in section 4. 
The remaining $\mu_3$-dependence of the CW-method can be used to 
measure the accuracy of
the calculation. We fix the scale $\mu_3$ so that the convergence 
of perturbation
theory is maximized near the minimum with a method
discussed in chapter \ref{scale}. We vary this fixed scale used in the 
calculation of $\langle\phi^{\ast}\phi(\mu_3)\rangle$ within limits
$0.5\mu_{\mbox{\scriptsize fix}} < \mu_3 < 2\mu_{\mbox{\scriptsize fix}}$. 
To compare these results with the Monte
Carlo data we must also run $\langle\phi^{\ast}\phi(\mu_3)\rangle$ to 
$\mu_3 = g_3^2$ using the running
\beq 
 \langle\phi^{\ast}\phi(\mu_1)\rangle -\langle\phi^{\ast}\phi(\mu_2)\rangle
 =\frac{g_3^2}{8\pi^2}\log \frac{\mu_1}{\mu_2}.
\eeq
Table~\ref{tbl} shows that the CW-values of the scalar condensate are almost 
independent of $\mu_3$ and thus more accurate than the results calculated
with $\hbar$ expansion.

\subsection{Propagators and particle masses}

\begin{figure}[t]
\vspace*{-3cm}
\hspace*{-3.75cm}
\epsfysize=8cm
\epsffile[0 550 300 800]{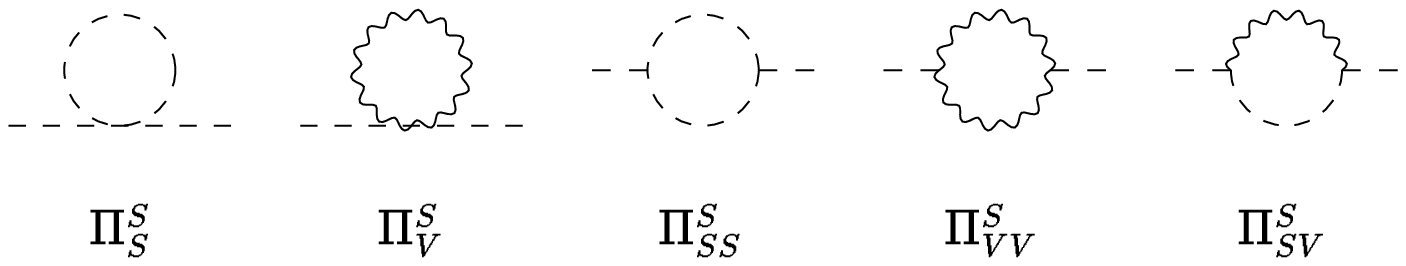}
\vspace*{-1cm}
  \caption{The 1-loop diagrams contributing to the scalar propagator.}
  \label{scalprop}
\end{figure}

\begin{figure}[t]
\vspace*{-3cm}
\hspace*{-3.75cm}
\epsfysize=8cm
\epsffile[0 550 300 800]{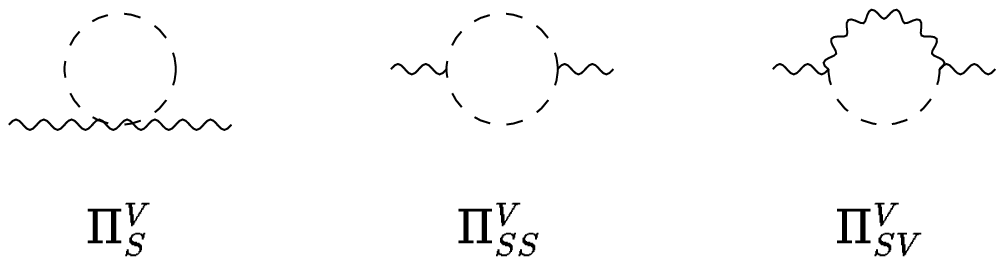}
\vspace*{-1cm}
  \caption{The 1-loop diagrams contributing to the vector propagator.}
  \label{vectprop}
\end{figure}

The perturbative masses of the system are derived as poles
of the corresponding propagators. Here we calculate the scalar and
vector propagator up to 1-loop level. The propagators are
\bea
    \langle \phi_1(-k)\phi_1(k)\rangle & = & \frac{1}{k^2+m_1^2-\Pi^S(k^2)},
\eea
\bea
    \langle A_i(-k)A_j(k)\rangle & = & \frac{\delta_{ij} 
-\frac{p_i p_j}{p^2}}{k^2+m_T^2-\Pi^V(k^2)} + \mbox{longitudinal}.
\eea
At 1-loop level the propagators contain diagrams in
Fig.~\ref{scalprop} and \ref{vectprop}.

The scalar self-energy is the sum of the following terms 
(Figure~\ref{scalprop})
\bea
   \Pi_S^S & = & \lambda_3\left[3 A_0(m_1^2)+ A_0(m_2^2)\right], \\
   \Pi_V^S & = & 2 g_3^2 A_0(m_T^2), \\
   \Pi_{SS}^S & = & 2 \lambda_3^2 \phi^2 \left[ 9 B_0(k^2;m_1^2,m_2^2)
                    + B_0(k^2;m_2^2,m_2^2)\right], \\
   \Pi_{VV}^S &=& \frac{g_3^2}{2 m_T^2} \left\{ B_0(k^2;m_T^2,m_T^2)
                    \left[k^4 + 4k^2 m_T^2 + 8m_T^4\right] \right. 
                    \nonumber \\ 
              & & \left. -2 B_0(k^2;m_T^2,0)\left[k^2+m_T^2\right]^2 \right. 
                  \nonumber \\
              & & \left. +B_0(k^2;0,0)k^4 -2 A_0(m_T)^2 m_T^2 \right\}, \\
   \Pi_{SV}^S &=& \frac{g_3^2}{m_T^2} \left\{ B_0(k^2;m_T^2,m_2^2) \left[
                  k^4+2 k^2(m_T^2+m_2^2)+(m_T^2-m_2^2)^2\right] \right. 
                  \nonumber \\ 
              & & \left. -B_0(k^2;m_2^2,0)\left[k^2+m_2^2\right]^2 \right.
                  \nonumber \\
              & & \left. -A_0(m_2^2)m_T^2 + A_0(m_T^2)\left[m_T^2-m_2^2-k^2
                  \right]\right\},
\eea
where we have used the integrals
\bea
 A_0(m^2) & = & \int \frac{d^dp}{(2\pi)^d}\frac{1}{p^2+m^2} = -\frac{m}{4\pi} 
 \\ B_0(k^2;m_1^2,m_2^2) & = & \int \frac{d^dp}{(2\pi)^d}\frac{1}{[p^2+m_1^2]
 [(p+k)^2+m_2^2]} \nonumber \\
 & = & \frac{i}{8 \pi (k^2)^{1/2}} \log \frac{m_1 + m_2 - 
 i(k^2)^{1/2}}{m_1 + m_2 + i(k^2)^{1/2}}.
\eea

The vector self-energy contains the following parts (Figure~\ref{vectprop})
\bea
  \Pi_S^V    &=& g_3^2\left[ A_0(m_1^2)+ A_0(m_2^2)\right], \\
  \Pi_{SS}^V &=& -\frac{g_3^2}{2} \left\{ B_0(k^2;m_1^2,m_2^2) \left[
                 k^2+2(m_1^2+m_2^2)+\frac{(m_1^2-m_2^2)^2}{k^2}\right] 
                 \right. \nonumber \\
             & & \left. \left[A_0(m_1^2)-A_0(m_2^2)\right] 
                 \frac{m_1^2-m_2^2}{k^2}  - A_0(m_1^2)- A_0(m_2^2)\right\}, \\ 
  \Pi_{SV}^V &=& g_3^2 \left\{ B_0(k^2;m_T^2,m_1^2) \left[
                 -k^2 - 2m_1^2 + 6m_T^2 - \frac{(m_1^2-m_T^2)^2}{k^2}
                 \right] \right. \nonumber \\ 
             & & \left. -B_0(k^2;m_1^2,0)\left[\frac{\left(k^2 + 
                 m_1^2\right)^2}{k^2}\right] \right. \nonumber \\
             & & \left. +A_0(m_T^2)\left[1 +\frac{m_1^2 -m_T^2}{k^2}\right] 
                 +A_0(m_1^2)\frac{m_T^2}{k^2}\right\}.
\eea
The mean field dependent masses are evaluated at the minimum of the
RG-improved 2-loop effective potential.

\section{Monte Carlo simulations}

The validity of perturbative results derived in previous section can
be studied by performing Monte Carlo simulations of 3d U(1)+Higgs
theory. Discretizing the continuum action (\ref{L3}) in the usual way one
gets 
\begin{eqnarray}
   S&=&\beta_G\sum_{\Box} \mbox{Re}(1-U_\Box) \nonumber \\
    & &-\beta_H\sum_{{x}, \mu} \mbox{Re}\left[
         \phi^{\ast}(x) U_{\mu}(x)\phi(x + \hat{\mu})\right] \nonumber \\ 
    & &+\sum_{{x}} \phi^{\ast}(x)\phi({x}) + 
         \beta_R\sum_{{x}} \left(1-\phi^{\ast}(x)\phi({x})\right)^2,
\end{eqnarray}
where $U_\mu(x)$ is a gauge field located on the links of the lattice
and $\phi(x)$ is a complex scalar field located on sites. $U_\Box$ is
the product of gauge fields around a plaquette. This action has three
coupling constants $\beta_G, \beta_H$ and $\beta_R$, 
giving us a three dimensional parameter space. The continuum limit
$a\to 0$ is described by the point $\beta_G=\infty, \beta_H=1/3,
\beta_R = 0$. 

\subsection{Constant physics curve}

To be able to compare the results from our simulations with the continuum
theory we have to establish the relation between lattice parameters
$\beta_G, \beta_H$ and $\beta_R$ and continuum parameters $x,y$ and
$g_3^2$. This is 
done by a constant physics curve, which can be calculated using
lattice perturbation theory. Due to the fact that the continuum
couplings are dimensionful, an exact relation can be found with a 2-loop
calculation \cite{ct}. The relevant relations are
\begin{eqnarray}
\beta_G &=& {1\over a g_3^2}, \nonumber  \\
\beta_R &=& {x \beta_H^2\over 4 \beta_G},  \nonumber \\
2 \beta_G^2\left({1\over\beta_H} - 3 - 
	 {2\beta_R\over \beta_H}\right) &=&
 y -(2+4x){\Sigma\beta_G\over 4 \pi}   \nonumber \\ &&-{1\over
16\pi^2}\bigg[\left(-4+8x-8x^2\right)  \nonumber \\ &&
\times \left(\ln 6\beta_G + 0.09\right) + 25.5 + 4.6x\bigg],
\end{eqnarray}
where $\Sigma=3.176$. We used $x=0.0463$ ,corresponding to $m_H=30$
GeV according to equation (\ref{aks}), in all simulations and 
varied the parameter $y$, which corresponds to temperature according
to equation (\ref{yy}), between $-1$ and $-0.025$. At the lower end of
this range perturbation theory is expected to work rather well, while 
approaching the critical point $y_c \simeq 0$ it should break down.

\subsection{Algorithms used}

By writing the Higgs field as $\phi(x) = R(x) e^{i\varphi(x)}$ and gauge field
as $U_\mu(x) = e^{i\theta_\mu(x)}$, we obtain the following expression for
action
\begin{eqnarray}
   S&=&\beta_G\sum_{x} \left(1-\cos\left(\theta_\Box\right)\right) \nonumber \\
    & &-\beta_H\sum_{{x}, \mu} R(x)\cos\left(-\varphi(x) + \theta_\mu(x) + 
         \varphi(x+\hat{\mu})\right)R(x+\hat{\mu}) \nonumber  \\
    & &+ \sum_{{x}} R(x)^2 + \beta_R\sum_{{x}} \left(1-R(x)^2\right)^2. 
\end{eqnarray}

We used an overrelaxed heatbath algorithm for the gauge field. The
heatbath algorithm used for the gauge field is described in
\cite{hn}. This local heat bath was then combined with standard
overrelaxation, where the angle is flipped to the other side of the minimum.

The form of the action would suggest separation of the radial and
angular modes of the Higgs field, in which case one could by change of
integration variables eliminate the angle of the Higgs field
as a dynamical variable, thus possibly saving some computer
time. However, the slow 
modes of the system are associated with the 
radius of the Higgs field and therefore it is essential to update
radial part as efficiently as possible. The obvious overrelaxation
algorithm \cite{fj} for the radial part of the Higgs field doesn't improve
the correlation time significantly. This is due to a rather poor 
acceptance ratio.

The best way to cope \cite{nonpert} with this problem seems to be not
to separate the 
radial and angular update of the Higgs field, but rather update the
cartesian components. This is done by defining two dimensional vectors
\begin{equation}
{\bf r} = \left( \begin{array}{c}
	R \cos\varphi \\
	R \sin\varphi
	\end{array} \right),
\end{equation}
and
\begin{equation}
{\bf v} = \left( \begin{array}{c}
	f\cos\theta \\
	f\sin\theta
	\end{array} \right).
\end{equation}
Now the local action
\[
S_{\mbox{\scriptsize loc}} = f R \cos(\phi + \phi_v) + c_1 R^2 + c_2 R^4,
\]
can be written as
\begin{equation}
S_{\mbox{\scriptsize loc}} = {\bf v \cdot r} + c_1 R^2 + c_2 R^4.
\end{equation}
If we now choose the components of the vector $\bf r$ perpendicular
and orthogonal to $\bf v$, denoted by $x$ and $y$, we get
\begin{equation}
S_{\mbox{\scriptsize loc}} = f x + c_1 (x^2 + y^2) + c_2 (x^2 + y^2)^2,
\label{sloc}
\end{equation}
which suggests the following overrelaxation algorithm: first we update
$y\to -y$, which corresponds to overrelaxation of the phase of the
Higgs field, and then we solve the equation (\ref{sloc}) to get a new
value for $x$ leaving the action invariant. 

For an exact overrelaxation algorithm one must leave the total path integral
invariant, in this case this means that if we change $x$ to $x'$ we
should have
\begin{equation}
{dS_{\mbox{\scriptsize loc}}(x)\over dx}^{-1} \exp(-S_{\mbox{\scriptsize loc}}(x)) = 
{dS_{\mbox{\scriptsize loc}}(x')\over dx'}^{-1} \exp(-S_{\mbox{\scriptsize loc}}(x')).
\end{equation}
Following \cite{nonpert} we now adopt an approximate method and
choose $x'$ so that the local action $S_{\mbox{\scriptsize loc}}$ is invariant and to
accept this change with probability
\begin{equation}
p = \min({dS_{\mbox{\scriptsize loc}}(x')\over dx'}/{dS_{\mbox{\scriptsize loc}}(x)\over dx}, 1).
\end{equation}
In our simulations the acceptance ratio was between 99.5 -
99.9\%. This is similar to values reported in three dimensional
simulations of SU(2)+Higgs theory \cite{nonpert}.

The simulations were done using Cray C90 at Center of Scientific
Computation in Helsinki. 

\subsection{Condensates}

We concentrated mainly on the scalar condensate, for which the continuum
and lattice normalizations are related by
\begin{equation}
\langle\phi^{\ast}\phi\rangle = {\beta_H\over 2 a}\langle R^2\rangle.
\end{equation} 
The 2-loop perturbative result calculated in $\overline{\rm MS}$ scheme
is given by equation (\ref{co}), which must be related to lattice
regularization scheme. One gets \cite{ct}
\begin{eqnarray}
\langle \phi^{\ast}\phi \rangle &=& 
             {1\over 2} \beta_H \beta_G g_3^2 \langle R^2 \rangle - 
             {g_3^2 \beta_G \Sigma \over 4 \pi} \nonumber  \\
				& &
             -{g_3^2\over 8 \pi^2}
               \left[ \log(6\beta_G)+\zeta+{\Sigma^2\over 4}-\delta\right],
\label{pdp}
\end{eqnarray}
where $\Sigma=3.176$, $\zeta=0.09$ and $\delta=1.94$. The scale
parameter $\mu$ has been chosen to be equal to $g_3^2$.

\begin{figure}
  \centering
  \leavevmode
  \epsfysize=7.5cm
  \epsfbox{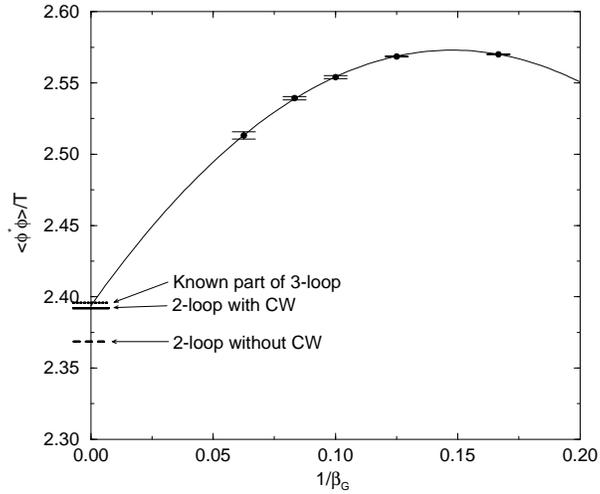}
  \caption{The continuum limit of scalar condensate $R^2$ at
$y=-1$. Also plotted are the perturbative results obtained both with
and without CW-optimization. The 3-loop results refers
to known part of unoptimized $\hbar$ expansion.}
  \label{r2cont1}
\end{figure}

\begin{figure}
  \centering
  \leavevmode
  \epsfysize=7.5cm
  \epsfbox{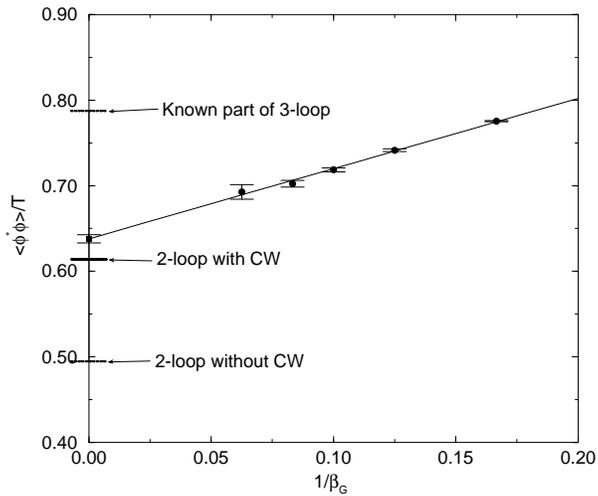}
  \caption{Same as Fig.~\protect\ref{r2cont1} but at $y=-0.025$.}
  \label{r2cont2}
\end{figure}

We can now compare Monte Carlo data with perturbative calculations
using equation (\ref{pdp}). To be able to do this we must extrapolate our 
results both to infinite volume and to zero lattice spacing. To
achieve the infinite volume limit we made sure that our results 
did not change with increasing lattice size. The actual lattices we
used were rather large, typically $60^3$. For the continuum limit
extrapolation we performed all our simulations with several values of
$\beta_G$. The leading order of 
finite lattice spacing effects is linear in lattice constant. However,
at larger lattice spacings we also observed higher order corrections
and as a consequence we did either a linear or a quadratic extrapolation
to continuum as appropriate. Examples of fits can be seen at Figures~\ref{r2cont1}-\ref{r2cont2}. 

\begin{figure}
  \centering
  \leavevmode
  \epsfysize=8cm
  \epsfbox{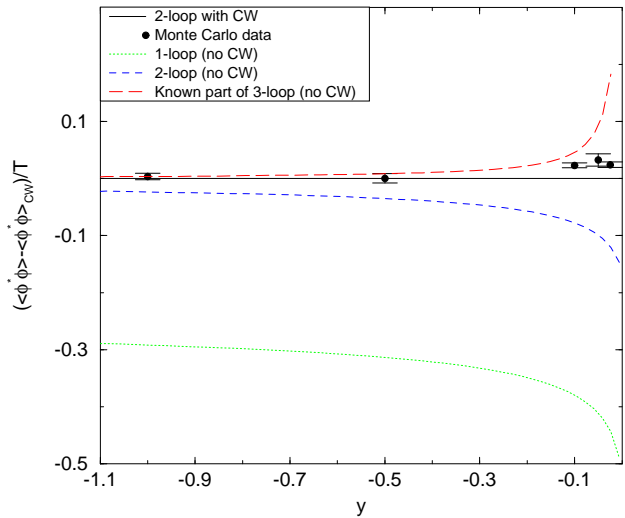}
  \caption{Perturbative results of the scalar condensate compared to Monte
Carlo data. The straight line is the CW-optimized result, which has been substracted from the others.}
  \label{r2comp}
\end{figure}

Ordinary perturbation theory quickly breaks down as one approaches the
critical value $y_c$ as can be seen from Figure~\ref{r2comp},
where we have 
plotted the Monte Carlo data with 1 and 2-loop straightforward
$\hbar$-perturbation 
theory results. Also plotted in the Figure~\ref{r2comp} are the the
known part of  
the 3-loop result and CW-improved result. 

The CW-method is in principle applicable even above the
critical temperature and seems to produce reasonable results at
$y$ values up to $y_c$. However, the CW-results are
consistently below the Monte Carlo data. This discrepancy is small (typically
less than 4\%) but statistically meaningful. The systematic error 
of CW method is not sufficient to explain this effect, as can be seen
from table~\ref{tbl}, so the effect is probably due to higher order
corrections. 

\begin{figure}
  \centering
  \leavevmode
  \epsfysize=7.5cm
  \epsfbox{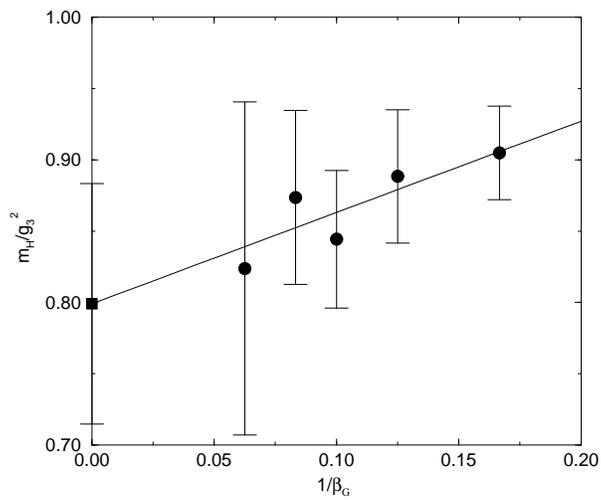}
  \caption{Continuum limit of scalar mass at $y=-0.05$.}
  \label{mHcont}
\end{figure}

\begin{figure}
  \centering
  \leavevmode
  \epsfysize=7.5cm
  \epsfbox{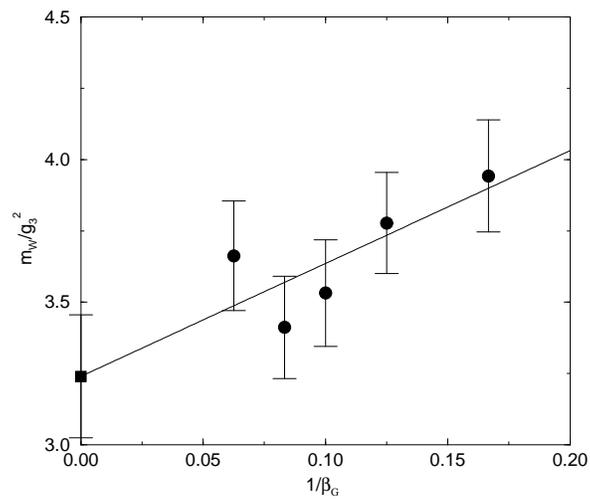}
  \caption{Continuum limit of vector mass at $y=-0.05$.}
  \label{mWcont}
\end{figure}

\begin{figure}
  \centering
  \leavevmode
  \epsfysize=7.5cm
  \epsfbox{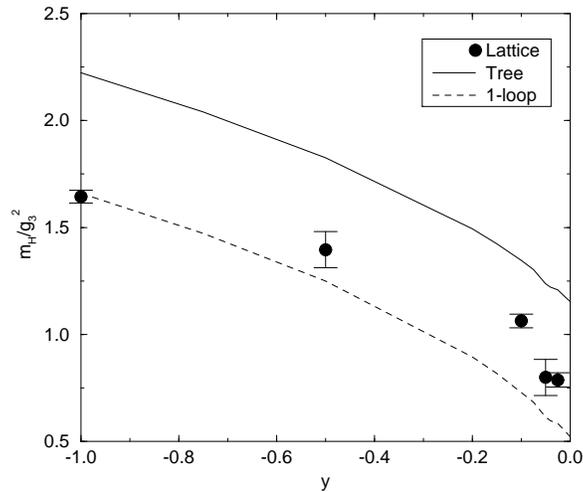}
  \caption{Perturbative results of scalar mass compared to Monte
Carlo data.}
  \label{mHcomp}
\end{figure}

\subsection{Masses}

The masses of scalar and vector particle at finite temperature were
measured from the exponential decay of suitably chosen
correlators. For scalar particle we used the operator
\[
\phi^{\ast}(x)\phi(x),
\]
and for vector
\[
\Im\phi^{\ast}(x)U_i(x)\phi(x+i).
\]

The quality of our data for masses is not as good as the quality of the scalar
condensate data, as can be seen from
Figures~\ref{mHcont}~and~\ref{mWcont}, where we have 
plotted the results of the continuum extrapolation of both scalar and
vector masses. However, the data is accurate enough to make some
conclusions on the validity of perturbation theory.

The tree level calculation and the 1-loop values of scalar mass are plotted in
Figure~\ref{mHcomp} with Monte Carlo data. The Monte Carlo data lies
between tree 
and 1-loop values and is not consistent with either of them, implying
that 1-loop calculation is probably not sufficient. 

\begin{figure}
  \centering
  \leavevmode
  \epsfysize=7.5cm
  \epsfbox{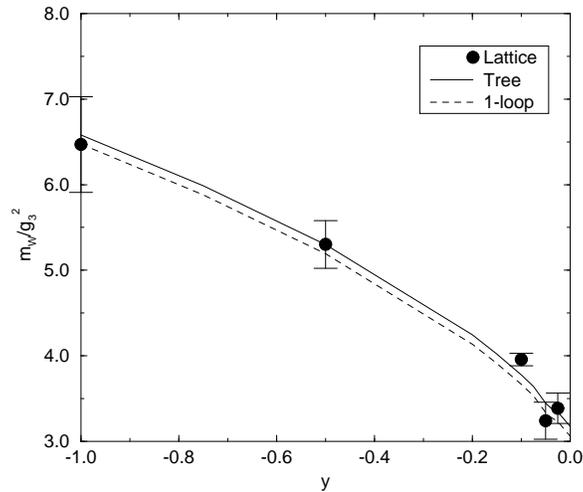}
  \caption{Perturbative results of vector mass compared to Monte
Carlo data.}
  \label{mWcomp}
\end{figure}

The Monte Carlo data for the vector mass is consistent with both 1-loop
and tree values, as can be seen from Figure~\ref{mWcomp}. However the
statistical errors are so large that it is not possible to make 
definite conclusions on the validity of the perturbation theory.

The results obtained here are rather similar to those found in
3d SU(2)+
Higgs 
theory \cite{nonpert}. There it was found that
the perturbative 1-loop calculation of vector mass agree extremely
well with Monte Carlo data. To be able to confirm this result in
U(1)+Higgs theory we would need to obtain more accurate Monte Carlo data 
for vector mass.

\section{Conclusions}

We have studied the validity of perturbation theory in the broken
phase of the U(1)+Higgs theory. Firstly, by measuring the scalar condensate
we examined
the improvement of perturbation theory when increasing the number of loops
and
also the effect of the CW optimization procedure. Our results show that
the 1-loop calculation is accurate only at $y$ values much lower than the
critical value $y_c$.  At $y=-1$, which is the smallest value of $y$
we examined, the error was already 13\% and at largest value
$y=-0.025$ the error was 73\%. Even the 2-loop calculation is not
accurate near $y_c$. The error was only 1\% at $y=-1$, but 22\% at
$y=-0.025$. However, optimized 
perturbation theory provides accurate estimates at 2-loop level even
up to the critical temperature. The largest error was 4\% at $y=-0.05$.
This indicates the importance of
optimized 2-loop perturbative results.

The 1-loop perturbative calculation of the scalar mass shows deviations 
from Monte Carlo data. The average difference between perturbative and
lattice results was 25\%. We interpret this as a result of the
insufficiency of a 1-loop perturbative calculation. This interpretation
is supported by the fact that the 1-loop values of scalar condensate
also show similar deviations. 
For the condensate 2-loop computations
are available and their inclusion improves the situation
considerably. For the scalar mass a 2-loop result is
unfortunately available only for the pure scalar case \cite{rajantie}. 
It would be very interesting to have the 2-loop result for
the scalar propagator also for gauge theories; then perturbation
theory could be optimized and probably a much better agreement
could be obtained.

The accuracy of our data for the vector mass has to improve before we can
make any definite conclusions on the validity of perturbation
theory for that quantity. One possibility would be to use the smearing
or fuzzying 
technique \cite{smear}, which has
proven to be effective in SU(2)+Higgs theory \cite{ptw} at
sufficiently high or low temperatures. 

\subsubsection*{Acknowledgements}

The authors thank K. Kajantie for his support and encouragement on
preparation of this paper. We also thank M. Laine for many discussions
and comments and
K. Rummukainen for his help on implementing the simulation program.

\end{document}